\documentclass[english,aps,prl,twocolumn,showpacs]{revtex4}
\usepackage[T1]{fontenc}
\usepackage[latin9]{inputenc}
\usepackage{color}
\usepackage{babel}
\usepackage{verbatim}
\usepackage{amstext}
\usepackage{amssymb}
\usepackage{graphicx}
\usepackage[unicode=true,pdfusetitle,
 bookmarks=false,
 breaklinks=false,pdfborder={0 0 1},backref=false,colorlinks=true]{hyperref}
\usepackage{breakurl}
\usepackage{upgreek}
\usepackage{epstopdf}
\makeatletter
\@ifundefined{textcolor}{}
{
 \definecolor{BLACK}{gray}{0}
 \definecolor{WHITE}{gray}{1}
 \definecolor{RED}{rgb}{1,0,0}
 \definecolor{GREEN}{rgb}{0,1,0}
 \definecolor{BLUE}{rgb}{0,0,1}
 \definecolor{CYAN}{cmyk}{1,0,0,0}
 \definecolor{MAGENTA}{cmyk}{0,1,0,0}
 \definecolor{YELLOW}{cmyk}{0,0,1,0}
 }
\makeatother
\begin{document}

\title{Scalable Implementation of Boson Sampling with Trapped Ions }
\author{C. Shen, Z. Zhang and L.-M. Duan}
\affiliation{Department of physics, University of Michigan, Ann Arbor, USA\\
and Center for Quantum Information, IIIS, Tsinghua University, China}
%\affiliation{}

\begin{abstract}
Boson sampling solves a classically intractable problem by sampling from a
probability distribution given by matrix permanents. We propose a scalable
implementation of Boson sampling using local transverse phonon modes of
trapped ions to encode the Bosons. The proposed scheme allows deterministic
preparation and high-efficiency readout of the Bosons in the Fock states and
universal mode mixing. With the state-of-the-art trapped ion technology, it
is feasible to realize Boson sampling with tens of Bosons by this scheme,
which would outperform the most powerful classical computers and constitute
an effective disproof of the famous extended Church-Turing thesis.
\end{abstract}

\maketitle

What is the ultimate computational power of physical devices? That is a deep
question of great importance for both physics and computer science. The
famous extended Church-Turing thesis (ECTT) postulates that a (classical)
probabilistic Turing machine can efficiently simulate the computational
power of any physical devices ("efficiently" here means with a polynomial
overhead) \cite{aaronson_2011}. The recent development in quantum
computation brings doubt to the ECTT with discovery of superfast quantum
algorithms. The most well known example is Shor's algorithm to factorize a
large number in polynomial time with a quantum computer \cite{Shor}.
Classically, whether factoring is hard is not settled (a "hard" problem
means its solution requires exponential time). No efficient classical
algorithm has been found yet to solve factoring, but it wouldn't be very
surprising if one finds one as this will not induce dramatic change to the
computational complexity theory.

Recently, Ref. \cite{aaronson_2011} introduces another problem, called Boson
sampling, which is hard for classical computers but can be solved
efficiently with a quantum machine. Boson sampling is defined as a problem
to predict the probabilities of the measurement outcomes in the Fock basis
for $M$ Bosonic modes, which start in definite Fock states and undergo a
series of mode mixing defined in general by a unitary matrix. By definition,
this problem can be efficiently solved with a quantum machine, but
classically its solution requires sampling of a probability distribution
given by matrix permanents with an exponentially large number of possible
outcomes. Computation of the matrix permanent is known to be $\#P$-hard
(much harder than the more well-known class of the NP-hard problems) \cite%
{valiant_1979}. Ref. \cite{aaronson_2011} rigorously proved that  Boson
sampling is classically intractable unless the so-called polynomial
hierarchy in the computational complexity theory collapses, which is
believed to be extremely unlikely. In this sense, compared with the
factoring problem, although Boson sampling has no immediate practical
applications, it is a problem much harder for classical computers to solve.
A demonstration of Boson sampling with a quantum machine
thus constitutes an effective disproof of the famous ECTT. Because of this
far-reaching theoretical implication, experimental demonstration of the
Boson sampling has raised strong interest recently. Several publications
this year have reported proof-of-principle demonstrations of the Boson
sampling with up to three photons \cite{e1,e2,e3,e4}. The key challenge for
the next-step experiments is to scale up the number of Bosons. The
demonstration using photons based on the spontaneous parametric down
conversion source has difficulty in terms of  scalability \cite%
{e1,e2,e3,e4}. The success probability decreases very rapidly with the
number of photons due to the probabilistic nature of the single-photon
source and the significant photon loss caused by the detector and the
coupling inefficiencies. This, in practice, limits the number of Bosons
below $10$, which is still within the simulation range of classical
computers.

In this paper, we propose a scalable scheme to realize Boson sampling using
the transverse phonon modes of trapped ions. Compared with the
implementation using photons, this scheme has the following desirable
features: First, the Fock states of the phonons can be prepared in a
deterministic fashion and there is no limitation to the number of Bosons
that one can realize with this system. We encode the Bosons using the local
transverse phonon modes \cite{TM}, and the state initialization can be done
through simple Doppler cooling and one step of the sideband cooling that
applies to any number of ions. Second, we find a technique to do projective
detection of the phonon numbers for all the ions through sequential spin
quantum jump measurements. This gives an implementation of number-resolving
phonon detectors with near perfect efficiency, much higher than the
efficiency of typical single-photon detectors. Finally, we prove that
universal coherent mixing of different phonon modes can be achieved through
a combination of the inherent Coulomb interaction and simple laser-induced
phase shifts of the ions. Through this scheme, it is feasible to realize
Boson sampling for tens of phonons with the state-of-the-art trapped ion
technology. This scale has gone beyond the simulation capability of any
classical computers and corresponds to the most interesting experimental
region for test of the ECTT \cite{aaronson_2011,AA}.

The problem of Boson sampling is defined as follows: we have $M$ input
Bosonic modes $a_{i}$ ($i=1,\,2,\,...,\,M$), which undergo coherent mode
mixing described in general by a unitary matrix $\Lambda $, with the output
modes given by $b_{i}=\sum_{j}^{M}\Lambda _{ij}a_{j}$. The input modes are
prepared in a Fock (number) state $\left\vert \mathbf{T}\right\rangle
=\left\vert t_{1},\,t_{2},\,...,\,t_{M}\right\rangle $, where $t_{i}$ is an
integer denoting the occupation number of the mode $a_{i}$. We measure the
output modes $b_{i}$ in the Fock basis and the probability to get the
outcome $\left\vert \mathbf{S}\right\rangle =\left\vert
s_{1},\,s_{2},\,...,\,s_{M}\right\rangle $ is given by \cite%
{scheel_2004,aaronson_2011}
\begin{equation}
P(\mathbf{S}|\mathbf{T})=\frac{\left\vert Per\left( \Lambda ^{(\mathbf{S},%
\mathbf{T})}\right) \right\vert ^{2}}{\prod_{j=1}^{M}s_{j}!\,%
\prod_{i}^{M}t_{i}!}  \label{eq:prob_ST}
\end{equation}%
where $Per(\cdot )$ denotes the matrix permanent and $\Lambda ^{(\mathbf{S},%
\mathbf{T})}$ is a sub-matrix of $\Lambda $ formed by taking $s_{j}$ copies
of the $j$-th column and $t_{i}$ copies of the $i$-th row of the matrix $%
\Lambda $. Since the total number of Bosons is conserved $%
N=\sum_{i}^{M}a_{i}^{\dagger }a_{i}=\sum_{j}^{M}b_{j}^{\dagger }b_{j}$, the
sub-matrix $\Lambda ^{(\mathbf{S},\mathbf{T})}$ has dimension $N\times N$.
Due to the hardness to calculate the matrix permanent, it becomes impossible
to sample the probability distribution $P(\mathbf{S}|\mathbf{T})$ with any
classical computer when the number of Bosons $N$ increases beyond $20\sim 30$%
. An experimental demonstration of a quantum machine that can successfully
perform this job therefore provides strong evidence against the ECTT.

To realize Boson sampling with trapped ions, we consider a chain of ions in
a linear Paul trap with the transverse trapping frequency $\omega _{x}$
significantly large than the axial one $\omega _{z}$. The Bosons are
represented by the local transverse phonon modes $a_{i}$ associated with
each ion $i$ ($i=1,\,2,\,...,\,M$), all with the oscillation frequency $%
\omega _{x}$. The Coulomb interaction between the ions introduces a small
perturbation to the oscillation frequency of the local phonon modes, with
the interaction Hamiltonian described by \cite{BHM_prl}
\begin{equation}
H_{c}=\sum_{1\leq i<j\leq M}\hbar t_{i,j}\left( a_{i}^{\dagger
}a_{j}+a_{i}a_{j}^{\dagger }\right) ,
\end{equation}%
where the hopping rates $t_{i,j}=t_{0}/\left\vert z_{i0}-z_{j0}\right\vert
^{3}$ and $t_{0}=e^{2}/\left( 8\pi \epsilon _{0}m\omega _{x}\right) $. Here,
$z_{i0}$ denotes the axial equilibrium position of the $i$th ion with mass $m
$ and charge $e$. The Hamiltonian (1) is valid under the condition $%
t_{i,j}\ll \omega _{x}$, which is always satisfied for the parameters
considered in this paper. To make the scheme more scalable and eliminate the
challenging requirement of resolving phonon sidebands for a large ion chain,
we use the local transverse phonon modes to represent the target Bosons
instead of the conventional normal modes.

{} To initialize the local phonon modes $a_{i}$ to the desired Fock states,
first we cool them to the ground state by laser cooling. The routine Doppler
cooling achieves a temperature $T_{D}\sim \hbar \Gamma /\left( 2k_{B}\right)
$ ($\Gamma $ is the natural bandwidth of the excited state and $k_{B}$ is
the Boltzmann constant), with the corresponding thermal phonon number $\bar{n%
}_{x}=k_{B}T_{D}/\hbar \omega _{z}\sim \Gamma /\left( 2\omega _{x}\right) $,
which is about $1\sim 2$ under typical values of $\omega _{x}\approx 2\pi
\times (5\sim 10)$ MHz and $\Gamma \approx 2\pi \times 20$ MHz. The sideband
cooling can further push the transverse modes to the ground state with $\bar{%
n}_{x}\approx 0$ \cite{SC}. For the axial modes, we only require their
thermal motion to be much less than the ion spacing, which is satisfied
already under routine Doppler cooling. As all the local transverse modes
have the same frequency (with $t_{i,j}\ll \omega _{x}$), we only need to
apply one step of the sideband cooling independent of the number of ions,
with the laser detuning set at $-\omega _{x}$. The off-resonant process in
the sideband cooling limits $\bar{n}_{x}\sim \gamma /\omega _{x}$, where $%
\gamma $ is the rate of the sideband cooling which needs to be comparable
with the phonon hopping rate $t_{i,i+1}$. For a harmonic trap, we take $%
l_{0}=\left[ e^{2}/\left( 4\pi \epsilon _{0}m\omega _{z}^{2}\right) \right]
^{1/3}$ as the length unit so that the ion spacings in this unit take
universal dimensionless values (of the order of $1$) independent of the ion
species and the trap frequency \cite{JAMES_ION_POSITION}. The hopping rate $%
t_{i,i+1}\sim t_{0}/l_{0}^{3}=\omega _{z}^{2}/\left( 2\omega _{x}\right) $
and the thermal phonon number after the sideband cooling $\bar{n}_{x}\sim
t_{i,i+1}/\omega _{x}\sim \omega _{z}^{2}/\left( 2\omega _{x}^{2}\right)
<10^{-2}$ with a typical $\omega _{z}\approx 2\pi \times \left( 0.3\sim
1\right) $ MHz. After cooling of all the transverse modes to the ground
state, we can then set them to any desired Fock states through a sequence of
laser pulses blue detuned at $\omega _{x}$ \cite{phonon_prep_detect}. Note
that the ion spacing is about or larger than $10$ $\mu m$ under our choice
of the parameters, and under such a spacing it is reasonable to assume
individual addressing of different ions with focused laser beams. The
focused beam can prepare different local modes $a_{i}$ to different Fock
states $\left\vert n_{i}\right\rangle $. For implementation of the Boson
sampling, without loss of generality we actually can choose $n_{i}=1$, which
requires only one pulse for preparation. To make the phonon hopping
negligible during the preparation step, the sideband Rabi frequency $\Omega $
needs to be large compared with the hopping rate $t_{i,i+1}\sim \omega
_{z}^{2}/\left( 2\omega _{x}\right) \sim 2\pi \times \left( 10\sim
100\right) $ kHz, which is easy to satisfy under typical laser power.

After the state initialization, we need to coherently mix different phonon
modes. The inherent Coulomb interaction described by the Hamiltonian (1)
serves this purpose, however, it is constantly on without a tuning knob and
we need to introduce additional control parameters to realize different
unitary transformations between the $M$ modes. To achieve this goal, we
introduce a simple operation which induces a controllable phase shift for
any local phonon mode at any desired time. A laser pulse with duration $t_{p}
$ and detuning $\delta $ to the sideband induces an additional Hamiltonian $%
H_{i}=\hbar \left( \Omega _{i}^{2}/\delta \right) a_{i}^{\dagger }a_{i}$ ($%
\Omega _{i}$ is the sideband Rabi frequency applied to the target ion $i$),
which gives a phase shift $U_{\phi _{i}}=e^{i\phi _{i}a_{i}^{\dagger }a_{i}}$
to the mode $a_{i}$ with $\phi _{i}=\Omega _{i}^{2}t_{p}/\delta $. We choose
$\Omega _{i}^{2}/\delta \gg t_{i,i+1}$ so that the pulse can be considered
to be instantaneous over the time scale of phonon tunneling.

The operation $U_{\phi _{i}}$ and the Coulomb interaction $H_{c}$ together
are universal in the sense that a combination of them can make any unitary
transformation on the $M$ phonon modes represented by the $M\times M$ matrix
$\Lambda $. Now we prove this statement. It is known that any unitary
transformation $\Lambda $ on $M$ Bosonic modes can be decomposed as a
sequence of neighboring beam-splitter-type of operations and individual
phase shifts \cite{Reck_universality}. The beam splitter operation for the
modes ($j,j+1$) is represented by the Hamiltonian $H_{bs}^{(j)}=\hbar
t_{j,j+1}\left( a_{j}a_{j+1}^{\dagger }+a_{j+1}a_{j}^{\dagger }\right) $. To
realize $H_{bs}^{(j)}$, we just need to cut off all the other interaction
terms in the Coulomb Hamiltonian given by Eq. (1) except for a specific pair (%
$j,j+1$). This can be achieved through the idea of dynamical decoupling
using the fast phase shifts $U_{\phi _{i}}$ with $\phi _{i}=\pi $ \cite%
{dynamical_decoupling}. Note that a Hamiltonian term $H_{ij}=\hbar
t_{i,j}\left( a_{i}a_{j}^{\dagger }+a_{j}a_{i}^{\dagger }\right) $ can be
effectively turned off for an evolution time $t$ if we apply an
instantaneous $\pi $-phase shift $U_{\phi _{i}=\pi }$ at time $t/2$ to the
mode $a_{i}$ to flip the sign of $H_{ij}$ to $-H_{ij}$ for the second half
period of the evolution. The interaction Hamiltonian $H_{c}$ has long-range
tunneling, but it decays fast with distance $d$ through $1/d^{3}$ scaling.
If we take the first order approximation to keep only the nearest neighbor
tunneling, the Hamiltonian has the form $H_{NN}=\sum_{i=1}^{M-1}\hbar
t_{i,i+1}\left( a_{i}^{\dagger }a_{i+1}+a_{i}a_{i+1}^{\dagger }\right) $.
The Hamiltonian $H_{NN}$ can be used to realize the required coupling $%
H_{bs}^{(j)}$ for an arbitrary $j$ if we apply $\pi $ phase shifts at time $%
t/2$ to every other modes in the ion chain except for the pair ($j,j+1$) as
illustrated in Fig. 1(a). This kind of decoupling can be extended and we can
simulate the Hamiltonian $H_{NN}$ (and thus $H_{bs}^{(j)}$) with the
original long range Hamiltonian $H_{c}$ to an arbitrary order of
approximation. Suppose we cut the interaction range in $H_{c}$ to the $k$th
order (i.e., we neglect the terms in $H_{c}$ that scale as $1/d_{ij}^{3}$
with $\left\vert i-j\right\vert >k$), we can shrink the interaction range
from $k$ to $k-1$ by applying one step of dynamical decoupling with the
pattern of $\pi $-phase shifts illustrated in Fig. 1(b). This step can be
continued until one reaches $H_{NN}$ through concatenation of the dynamical
decoupling \cite{dynamical_decoupling}. This proves that the Coulomb
interaction Hamiltonian $H_{c}$, together with the phase shifts $U_{\phi
_{i}}$ on single ions, can realize any beam splitter operations and thus be
universal for construction of arbitrary unitary operations on the $M$
phononic modes. We should note that the above proof of universality based on
the idea of dynamical decoupling is constructive but does not necessarily give
the optimal decomposition of a given unitary operation $\Lambda $. Direct
optimization of a sequence of control parameters $\phi _{i}$ in $U_{\phi
_{i}}$ and the evolution time could give much more efficient construction of
a given unitary matrix.

\begin{figure}[tbp]
\includegraphics[width=0.45\textwidth]{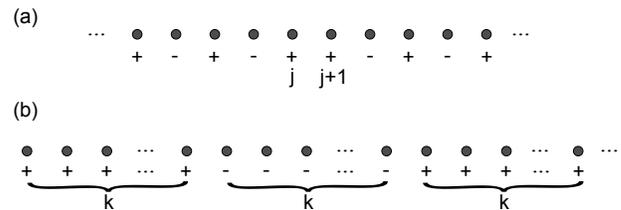}
\par
\label{fig:decoupling}
\caption{Control of the tunneling Hamiltonian through the dynamical decoupling.
The negative signs in (a,b) denote the set of ions to be
applied a $\protect\pi $ phase shift at half of the evolution time while the positive signs denote the ions
left intact. (a) The $\protect\pi $-phase pattern to turn off other tunneling terms in $H_{NN}$ 
except for a neighboring pair $j,j+1$; (b) The $\protect\pi $-phase pattern to shrink the tunneling range 
of the Hamiltonian from $k$ to $k-1$.}
\end{figure}

The final step of the Boson sampling is detection of all the phononic modes
in the Fock basis. The conventional method of measuring the phonon number
distribution of \ a single mode by recording the spin oscillation from red
or blue sideband pulses is not applicable here as it cannot measure
correlation of different phonon modes in the Fock basis \cite%
{phonon_prep_detect}. What we need is a projective measurement of each mode
in the Fock basis which gives information of arbitrary high order
correlations between different modes. For trapped ions, a projective
measurement of its spin (internal) state can be done with a very high
efficiency through the quantum jump technique using a cycling transition.
However, the spin detection gives only binary measurement outcomes ("dark"
or "bright"). We need to figure out a way to perform projective measurements
of the Fock states (with multiple possible outcomes) for each phonon mode
through the binary spin detection. This is achieved through a consecutive
detection scheme with the following steps: (1) First, to illustrate the
idea, we consider a single ion with its phonon mode in an arbitrary state $%
\sum_{n}c_{n}\left\vert n\right\rangle $ and its spin prepared in the dark
state $\left\vert D\right\rangle $ (see Fig. 2a). (2) Through the well known
adiabatic transition technique \cite{adia_manipu_motion}, we make a complete
population transfer from $\left\vert n+1\right\rangle \left\vert
D\right\rangle $ to $\left\vert n\right\rangle \left\vert B\right\rangle $
for all the Fock components $\left\vert n\right\rangle $ by chirping the
frequency of a laser pulse across the red detuning at $-\omega _{x}$ (see
Fig. 2b for the population distribution after this step). (3) We make a
carrier transition $\left\vert n\right\rangle \left\vert D\right\rangle
\rightleftarrows \left\vert n\right\rangle \left\vert B\right\rangle $ with
a $\pi -$pulse to flip the dark and the bright states (see Fig. 2c). (4)
After this step, we immediately measure the spin state of the ion through
the quantum jump detection. With probability $\left\vert c_{0}\right\vert
^{2}$, the outcome is "bright". In this case the measurement is finished and
we know the phonon is in the $\left\vert n=0\right\rangle $ state.
Otherwise, the spin is in the dark state and the phonon is in the $%
\left\vert n\geq 1\right\rangle $ components (see Fig. 2d for the population
distribution in this case). When the spin is in the dark state, the ion does
not scatter any photons during the quantum jump measurement. So its phonon
state will not be influenced by this measurement. This feature is important
for this consecutive measurement scheme. (5) Now with the phonons in the $%
\left\vert n\geq 1\right\rangle $ components, we just repeat the steps
(2)-(3)-(4) until finally we get the outcome "bright" for the spin
detection. We conclude that the phonon is in the Fock state $\left\vert
n=l\right\rangle $ if the outcome "bright" occurs (with probability $%
\left\vert c_{l}\right\vert ^{2}$) after $l$ repetitions of the above steps.
(6) The above consecutive measurement scheme can be extended
straightforwardly to measure $M$ local phonon modes in the Fock basis
independently with $M$ ions. The only requirement is that the phonon
tunneling between different modes is negligible during the measurement
process. The slowest step of the measurement is the quantum jump detection
of the ion spin state. Recently, there has been experimental report of high
efficiency ($>99\%$) spin state detection within $10\text{ }\mu \text{s}$
detection time \cite{fast_detection}. The typical phonon hopping rate
between the neighboring ions in our scheme is in the range of $t_{i,i+1}\sim
2\pi \times \left( 10\sim 100\right) $ kHz, and this hopping rate can be
significantly reduced during the detection through either an expansion of
the ion chain along the $z$ direction right before the direction by lowering
the axial trap frequency or application of a few dynamical decoupling pulses
to turn off the neighboring tunneling during the detection. As the hopping
scales as $1/d^{3}$, a moderate increase of the effective distance $d$ will
significantly reduce the tunneling and push it below the kHz level. We
should note that for the Boson sampling algorithm, the output phonon number
per mode is typically small (the conventional photon detectors actually can
only distinguish $0$ and $1$ photons), and the number of repetitions in our
consecutive measurement scheme is either zero or very few in most cases.

\begin{figure*}[tbp]
\includegraphics{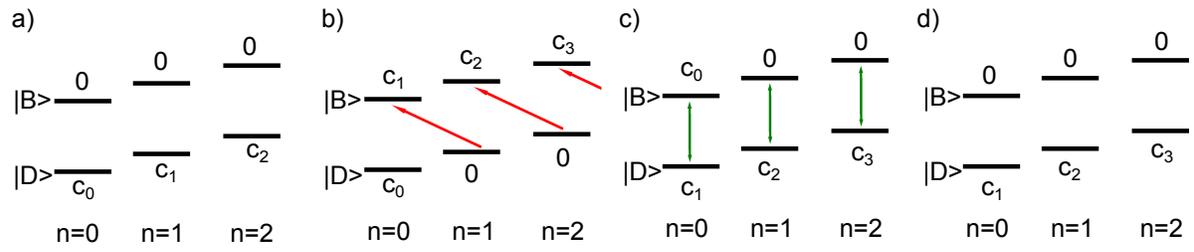}
\caption{A consecutive measurement scheme to perform projective detection of the
phonon mode in the Fock basis. (a) The initial state configuration right before the 
measurement.  (b-d) The state configuration after the blue sideband transition, the carrier
transition, and the quantum jump detection. These three steps are repeated until one 
finally registers the "bright" state (see the text for details). }
\label{fig:state_detection}
\end{figure*}

In summary, we have proposed a scalable scheme to realize the Boson sampling
algorithm by use of the local transverse phonon modes of trapped ions. The
scheme allows deterministic preparation and high-efficiency readout of the
phonon Fock states and universal manipulation of the phonon modes through a
combination of inherent Coulomb interaction and individual phase shifts.
Several dozens of ions have been successfully trapped experimentally to form
a linear chain, and in principle there is no limitation to the number of
ions that can be manipulated in a linear Paul trap by use of anharmonic
axial potentials \cite{Lin_anharmonic}. This scheme thus opens the
perspective to realize Boson sampling for dozens of phonons with the
state-of-the-art trapped ion technology, which would beat the capability of
any classical computers and give the first serious experimental test of the
extended Church-Turing thesis.

We thank Kihwan Kim for discussions. This work was
supported by the NBRPC (973 Program) 2011CBA00300 (2011CBA00302), the IARPA
MUSIQC program, the ARO and the AFOSR MURI programs, and the DARPA OLE
program.

\end{document}